\documentclass[runningheads]{llncs}
\usepackage{graphicx}
\usepackage{amsmath}
\usepackage{hyperref}
\begin{document}

\title{
VTutor: An Animated Pedagogical Agent SDK that Provide Real Time Multi-Model Feedback
}

\titlerunning{VTutor}
%
\author{Eason Chen\inst{1} \and
Chenyu Lin\inst{2}\and
Yu-Kai Huang\inst{1}
Xinyi Tang\inst{1}
Aprille Xi\inst{1}
Jionghao Lin\inst{1,3}
Kenneth Koedinger\inst{1}
}
\authorrunning{Eason Chen et al.}
%
\institute{
    Carnegie Mellon University \and
    New York University \and
    The University of Hong Kong
}

%
\maketitle              

\begin{abstract}
Pedagogical Agents (PAs) show significant potential for boosting student engagement and learning outcomes by providing adaptive, on-demand support in educational contexts. However, existing PA solutions are often hampered by pre-scripted dialogue, unnatural animations, uncanny visual realism, and high development costs. 
To address these gaps, we introduce VTutor, an open-source SDK leveraging lightweight WebGL, Unity, and JavaScript frameworks. VTutor receives text outputs from a large language model (LLM), converts them into audio via text-to-speech, and then renders a real-time, lip-synced pedagogical agent (PA) for immediate, large-scale deployment on web-based learning platforms. By providing on-demand, personalized feedback, VTutor strengthens students’ motivation and deepens their engagement with instructional material. 
Using an anime-like aesthetic, VTutor alleviates the uncanny valley effect, allowing learners to engage with expressive yet comfortably stylized characters. 
Our evaluation with 50 participants revealed that VTutor significantly outperforms the existing talking-head approaches (e.g., SadTalker) on perceived synchronization accuracy, naturalness, emotional expressiveness, and overall preference. 
As an open-source project, VTutor welcomes community-driven contributions—from novel character designs to specialized showcases of pedagogical agent applications—that fuel ongoing innovation in AI-enhanced education. By providing an accessible, customizable, and learner-centered PA solution, VTutor aims to elevate human-AI interaction experience in education fields, ultimately broadening the impact of AI in learning contexts. The demo link to VTutor is at \url{https://vtutor-aied25.vercel.app}.
\end{abstract}

\section{Introduction and Background}

AI in Education (AIED) aims to enhance learning processes and outcomes by leveraging artificial intelligence techniques \cite{self2016birth} such as intelligent tutoring systems \cite{stamper2024enhancing}, adaptive assessments \cite{minn2022ai}, programming \cite{chen2023gptutor}, and chatbot \cite{chen2022effect}. While large language models (LLMs) now excel at generating personalized content and responding to learners' open-ended queries \cite{chen2024gptutor,chen2024systematic,wang2024large,ali2024effects,kasneci2023chatgpt}, many of these systems are constrained by purely text-based or voice-based interfaces. \emph{Pedagogical Agents} (PAs) offers an opportunity for a more human-like interaction experience with LLMs \cite{sikstrom2024pedagogical,Lin2024,heeyoHeeyo}, delivering immediate, adaptive guidance via facial expressions, gestures, and voice \cite{Domagk2010,Veletsianos2012,Makransky2018,Davis2021}.

However, existing PA frameworks often rely on scripted interactions \cite{okado2023can,cramifyCramifyCram}, utilize static or non-synchronized facial expressions \cite{heeyoHeeyo}, high development cost, and long latency \cite{Lin2024,zhang2023sadtalker}. These constraints lead to interaction delays, unnatural facial movements, and overall lower engagement, thereby limiting the development of dynamic, learner-centered AI tutoring experiences. As a result, these PAs cannot reliably provide responsive feedback, emotional support, and adaptive scaffolding—key features that enhance the tutoring process. Moreover, when a PA design becomes overly humanlike, it risks triggering the ``uncanny valley'' phenomenon \cite{ciechanowski2019shades,sikstrom2024pedagogical}, wherein users feel discomfort toward entities that appear almost, but not exactly, human. This effect can discourage learners from interacting with the PA, undercutting its potential to facilitate meaningful learning outcomes. Consequently, there is a growing need for an integrated solution that aligns robust LLM-driven dialogue capabilities with visually appealing, synchronized, and scalable PA tools, all while avoiding the uncanny valley effect to preserve learner engagement and trust.

Meanwhile, virtual avatars have increasingly become a presence in real-world applications, particularly in the entertainment field, where many content creators opt to engage with their audiences using digital personas instead of revealing their real faces. These creators, known as \textit{VTubers} \cite{ferreira2022vtuber,chinchilla2024vtuber}, leverage face tracking and motion capture technologies to animate their avatars in real-time by 3D engine like Unity, enabling expressive and interactive performances. The aesthetic of VTuber avatars is heavily influenced by anime culture, incorporating cuteness and charm elements that make them visually appealing and emotionally engaging for viewers \cite{ferreira2022vtuber,chinchilla2024vtuber}. These stylistic choices not only enhance relatability but also circumvent the uncanny valley effect, maintaining user comfort and engagement.

\subsubsection{Our Contribution:} This paper introduces \textbf{VTutor}, an open-source Software Development Kit (SDK) designed to streamline the development and deployment of animated pedagogical agents (APAs) in educational settings. VTutor enhances human-AI interaction by enabling real-time, engaging, and emotionally expressive AI-driven tutors. Specifically, VTutor contributes:

\begin{enumerate}
    \item \textbf{A Generative AI-Driven Interactive Agent:} By integrating output from LLMs with animated pedagogical agents, VTutor facilitates real-time, personalized, and adaptive interaction experiences.
    \item \textbf{Customizable Characters:} VTutor adopts anime-style characters that strike a balance between realism and abstraction, avoiding the uncanny valley while enhancing engagement across diverse learners.
    \item \textbf{A Seamless, Scalable, and Developer-Friendly SDK:} The lightweight browser-compatible architecture ensures that VTutor can be effortlessly embedded into existing educational platforms with minimal setup.
\end{enumerate}

By addressing key challenges in traditional APAs—such as lack of real-time adaptability, poor lip synchronization, and high computational overhead—VTutor paves the way for more immersive and accessible AI-driven educational tools.


\section{VTutor System Implementation}

\begin{figure}[b!]
    \centering
    \includegraphics[width=0.5\linewidth]{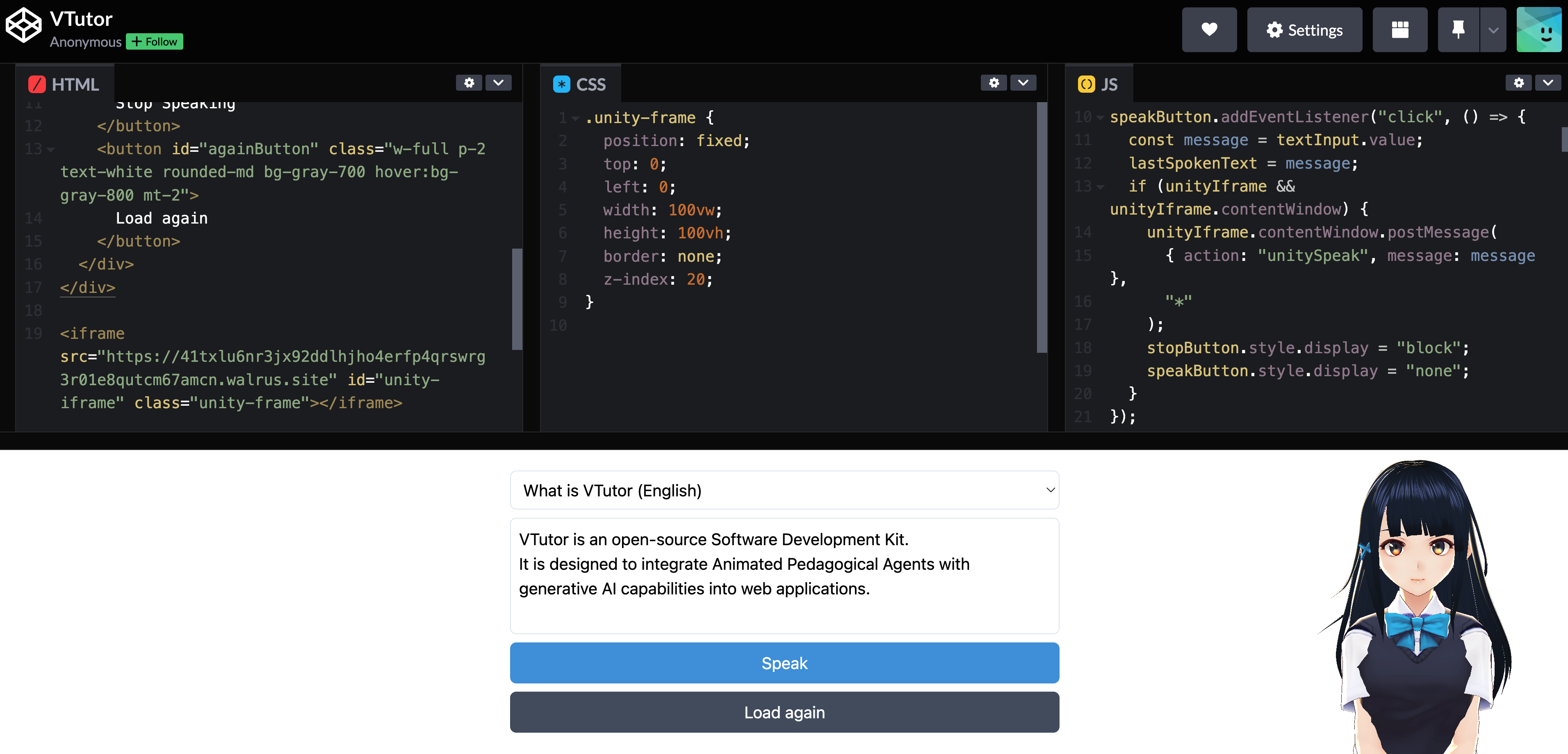}
    \caption{Screenshot of the VTutor SDK demo at \url{https://vtutor-aied25.vercel.app/sdk}, the few lines of codes in the image are what developers need to do to add VTutor to their own website. VTutor is the animated avatar presented at the bottom right.}
    \label{fig:enter-label}
\end{figure}

\vspace{-9pt}
\noindent {VTutor is composed of the following components:}
\vspace{-9pt}
\subsubsection{\textbf{(1) Large Language Model Integration.}}
VTutor presents AI-generated text responses by integrating various LLMs (e.g., GPT, Claude, Deepseek). Educators can customize prompts or knowledge bases to align PA interactions with learning objectives, enabling flexible, contextually adaptive dialogues.

\subsubsection{\textbf{(2) Text-to-Speech (TTS).}}
VTutor accepts audio in \texttt{.wav} format from major TTS services (e.g., Azure, ElevenLabs). This enables high-quality, natural speech that resonates with learners. During run-time, the system retrieves newly generated text from the LLM, converts it to audio via a TTS API, and sends it to the Unity engine for real-time facial expression and mouth rendering.

\subsubsection{\textbf{(3) LipSync Module.}}
A Unity lip-sync library \cite{githubGitHubHecomiuLipSync} analyzes incoming audio to detect phoneme sequences. Corresponding mouth-blend shapes are applied to a 3D or 2D avatar, achieving near real-time alignment with an average latency of under one second for short utterances. This approach yields more accurate and smooth mouth movements compared to naive keyframe or jaw-only animations.

\subsubsection{\textbf{(4) WebGL-Based Rendering.}}
VTutor packages the Unity environment into a WebGL build, allowing easy embedding in any web page by iframe. Communication between the web page and the embedded PA occurs via an open-sourced software development kit (SDK) library, enabling dynamic control and fine-grained interactivity (e.g., changing gestures, expressions, or switching avatars).

\section{Evaluation Method and Results}


We invited participants to compare VTutor with talking-head pedagogical agents generated using SadTalker \cite{zhang2023sadtalker}, following the methodology described by \cite{Lin2024}. To minimize potential appearance biases, we used the same VTutor agent design to create the SadTalker talking head. The full evaluation materials can be found at \url{https://vtutor-aied25.vercel.app/evaluation}.

Our experiment received approval from the Institutional Review Board (IRB). We recruited 50 participants for evaluation through convenience sampling via Prolific. The participants were between 20 and 50 years old, consisting of 30 males and 20 females. We conducted the experiment online, where participants watched the videos on YouTube and responded to the questions using Google Forms.

The videos featured VTutor and Sad Talker speaking, but throughout the entire experiment, they were referred to only as Agent A and Agent B. Among them, 25 random participants viewed VTutor on the left as Agent A and SadTalker on the right as Agent B, while the remaining 25 participants saw the opposite arrangement. After watching the videos, participants were asked which agent they preferred and were asked to rate the following dimensions for each agent on a scale from 1 to 7, and answered open-ended questions about their opinions.

\begin{itemize}
    \item \textbf{General Preference Score}: Preference rating between the two agents.
    \item \textbf{Sync Accuracy}: Evaluates the alignment between facial expressions, lip movements, and audio.
    \item \textbf{Naturalness}: Assesses whether the video appears natural and not rigid.
    \item \textbf{Emotional Expression}: Determines whether the facial expressions effectively convey the agent's emotions.
    \item \textbf{Visual Coherence}: Examines the coordination between lip movements and other facial features.
\end{itemize}

For quantitative data analysis, we used the Chi-square test for user preference. For the user rating, we conducted an independent samples t-test. We posed open-ended questions about the two agents to gather qualitative insights through thematic analysis \cite{fereday2006demonstrating}, which is well-suited for exploratory studies without the need of agreement scores as suggested by \cite{mcdonald2019reliability}, as our main goal was to capture a broad range of perspectives rather than measure inter-rater reliability.

\subsection{Quantitative User Preference Evaluation Results}


\begin{table}[h!]
\scriptsize
    \caption{Comparison of VTutor and SadTalker on different evaluation metrics}
    \centering
    \begin{tabular}{p{3.5cm}p{1.7cm}p{1.7cm}ccc}
        \hline
        \textbf{Question} & \textbf{VTutor Mean (SD)} & \textbf{SadTalker Mean (SD)} & \textbf{t statistics} & \textbf{p-value} & \textbf{Cohen's d} \\
        \hline
        General Preference Score & 5.00 (1.75) & 3.62 (1.51) & 4.2218 & 0.0001 & 0.8444 \\
        Sync Accuracy & 4.58 (1.82) & 3.66 (1.49) & 2.7642 & 0.0068 & 0.5528 \\
        Naturalness & 4.12 (2.04) & 3.10 (1.67) & 2.7390 & 0.0073 & 0.5478 \\
        Emotional Expression & 4.38 (1.94) & 3.30 (1.58) & 3.0546 & 0.0029 & 0.6109 \\
        Visual Coherence & 4.56 (1.64) & 3.64 (1.63) & 2.8142 & 0.0059 & 0.5628 \\
        \hline
    \end{tabular}

    \label{tab:comparison}
\end{table}

The participants exhibited a strong preference for the VTutor version. Among the 50 participants, 36 preferred VTutor, while only 14 favored Sad Talker. The statistical analysis yielded a chi-square statistic of $\chi^2(1, N = 50) = 9.68$, with a p-value of $p = .0019$ and Cramer's V of $V = .44$, indicating a moderate to strong effect size. Moreover, as shown in Table \ref{tab:comparison}, VTutor demonstrated a significant advantage across all evaluation dimensions ($p < .01$). These results suggest that VTutor, utilizing the Unity technology framework, provides a more engaging, synchronized, natural, emotionally expressive, and visually coherent PA compared to the Sad Talker-generated talking head.

\subsection{Qualitative Insights}

\subsubsection{\textbf{Concerns About SadTalker's Uncanny Valley.}}
Two participants explicitly mentioned being disturbed by SadTalker’s appearance. One stated, 
\emph{``[SadTalker] was a bit unsettling and looked strange''} (P35), 
while another directly commented, 
\emph{``[SadTalker] look weird and uncanny''} (P36).
In total, many participants (e.g., P24, P28, P35, P41, P45) described SadTalker as ``unsettling,'' ``creepy,'' or ``weird,'' suggesting an uncomfortable, uncanny-valley-like experience.

\subsubsection{\textbf{Positive Feedback and Remaining Issues on VTutor.}}
By contrast, VTutor received favorable remarks from several participants, particularly for its clarity and color. One noted, 
\emph{``[VTutor] was overall brighter and easier to see. Their lip movements also did a better job of matching the audio.''} (P50).
Meanwhile, another participant appreciated how its stylized mouth avoided some uncanny pitfalls: 
\emph{``It didn't look `natural,' but [VTutor] was a lot closer to an anime character in the mouth. That in itself made it feel less creepy.'} (P41).
Despite these positives, a few participants still felt VTutor’s facial expression animations were also ``weird'' or ``unnatural'' (P28, P40, P45). While its anime-like appearance seemed to alleviate the uncanny valley, further refinements on VTutor avatar are needed to achieve a more convincing, natural-looking facial expression.

\subsection{Computational Time and Latency}


In our experiment, SadTalker required 143 seconds to generate a 7-second video on an NVIDIA Titan XP GPU, whereas VTutor produced the animation within one second via a browser environment. This efficiency is critical for real-time interactive pedagogical agents. Moreover, VTutor can be integrated with just an HTML iframe and requires no specialized backend to compute the animation rendering, underscoring its lightweight and scalability in deploying APA at scale.

\section{Discussion}


\subsubsection{\textbf{Advancing Animated Pedagogical Agents.}}
VTutor redefines the design and capabilities of animated Pedagogical Agents (PAs) by leveraging generative AI and real-time animation. Existing solutions often suffer from either scripted dialogues or rigid, non-synchronized facial expressions. In contrast, VTutor’s modular pipeline integrates large language models (LLMs), text-to-speech (TTS), and real-time lip synchronization to provide personalized, adaptive feedback in a more fluid and engaging manner. This dynamic synergy ensures that the agent’s speech, mouth movements, and expressions are meticulously aligned, mitigating the robotic feel or delay that commonly characterizes earlier systems.

Beyond technical innovation, VTutor prioritizes the user experience. Our evaluation demonstrates that participants perceive the system to be more natural, well-synchronized, and visually appealing, which will foster deeper engagement. By delivering human-like, contextually responsive guidance, VTutor bolsters the motivational and emotional underpinnings crucial for effective learning.

\subsubsection{\textbf{Anime-Style Avatars and the Uncanny Valley.}}
A key insight emerging from our user study is that anime-like avatars may reduce the uncanny valley effect often encountered with realistic character models. Participants frequently described SadTalker's talking-head approaches as ``unsettling'' or ``creepy,'' highlighting how slight deviations from human realism can evoke discomfort. In contrast, VTutor’s animated style characters, with simpler facial features and intentional abstraction, were praised for feeling more natural and less uncanny. 

This suggests that designing PAs with an animated style is not merely an artistic choice but a strategic one to maintain user engagement. By balancing expressive detail with clear abstraction, anime-like avatars avoid the pitfalls of near-realistic facial features that sometimes trigger negative emotional responses. This finding has broad implications for PA design, emphasizing the value of recognizable yet stylized models that preserve learners’ comfort and trust.


\subsubsection{\textbf{Limitations and Future Directions.}}
Despite VTutor’s promising results, VTutor's animations still have room to improve. Incorporating refined 3D models or advanced animations could help produce more lifelike interactions. Additionally, VTutor’s current Unity WebGL build of approximately 115~MB can result in long initial loading times. Future efforts will focus on package optimization and progressive loading to enhance performance and scalability.

\section{Conclusion and Potential Applications}

VTutor seamlessly combines large language models (LLMs), text-to-speech, and lip-sync to render adaptive, anime-style pedagogical agents in a lightweight browser environment. Evaluation results show that VTutor outperforms conventional talking-head approaches in lip synchronization, naturalness, and emotional expressiveness. By offering an open-source, customizable toolkit, VTutor paves the way for more engaging AI-driven pedagogical agents interaction experiences.

Leveraging its user-friendly, real-time, and scalable architecture, VTutor could provide immediate and adaptive feedback by pedagogical agents using the output from LLMs and thus effectively addresses a wide range of educational needs. From high-impact STEM tutoring and language learning peers to personalized workforce skill development and learner social support, VTutor has the potential to enrich human-AI interactions across various educational domains.


\bibliographystyle{splncs04}
\bibliography{main}

\begin{thebibliography}{10}
\providecommand{\url}[1]{\texttt{#1}}
\providecommand{\urlprefix}{URL }
\providecommand{\doi}[1]{https://doi.org/#1}

\bibitem{ali2024effects}
Ali, O., Murray, P.A., Momin, M., Dwivedi, Y.K., Malik, T.: The effects of artificial intelligence applications in educational settings: Challenges and strategies. Technological Forecasting and Social Change  \textbf{199},  123076 (2024)

\bibitem{chen2022effect}
Chen, E.: The effect of multiple replies for natural language generation chatbots. In: CHI Conference on Human Factors in Computing Systems Extended Abstracts. pp. 1--F--5 (2022)

\bibitem{chen2023gptutor}
Chen, E., Huang, R., Chen, H.S., Tseng, Y.H., Li, L.Y.: Gptutor: a chatgpt-powered programming tool for code explanation. In: International Conference on Artificial Intelligence in Education. pp. 321--327. Springer (2023)

\bibitem{chen2024gptutor}
Chen, E., Lee, J.E., Lin, J., Koedinger, K.: Gptutor: Great personalized tutor with large language models for personalized learning content generation. In: Proceedings of the Eleventh ACM Conference on Learning@ Scale. pp. 539--541 (2024)

\bibitem{chen2024systematic}
Chen, E., Wang, D., Xu, L., Cao, C., Fang, X., Lin, J.: A systematic review on prompt engineering in large language models for k-12 stem education. arXiv preprint arXiv:2410.11123  (2024)

\bibitem{chinchilla2024vtuber}
Chinchilla, P., Kim, J.: Vtuber for streamers: Exploring the role of social presence in the visual representation of streamers. Communication Studies pp. 1--17 (2024)

\bibitem{ciechanowski2019shades}
Ciechanowski, L., Przegalinska, A., Magnuski, M., Gloor, P.: In the shades of the uncanny valley: An experimental study of human--chatbot interaction. Future Generation Computer Systems  \textbf{92},  539--548 (2019)

\bibitem{cramifyCramifyCram}
Cramify: {C}ramify: {T}he {N}ew {W}ay to {C}ram | {A}{I} {S}tudy {G}uides \& {M}ore --- cramify.ai. \url{https://cramify.ai/} (2025), [Accessed 23-01-2025]

\bibitem{Davis2021}
Davis, R.O., Park, T., Vincent, J.: A systematic narrative review of agent persona on learning outcomes and design variables to enhance personification. Journal of Research on Technology in Education  \textbf{53}(1),  89--106 (2021). \doi{10.1080/15391523.2020.1830894}

\bibitem{Domagk2010}
Domagk, S.: Do pedagogical agents facilitate learner motivation and learning outcomes?: The role of the appeal of agent’s appearance and voice. Journal of Media Psychology  \textbf{22}(2),  84--97 (2010). \doi{10.1027/1864-1105/a000011}

\bibitem{fereday2006demonstrating}
Fereday, J., Muir-Cochrane, E.: Demonstrating rigor using thematic analysis: A hybrid approach of inductive and deductive coding and theme development. International journal of qualitative methods  \textbf{5}(1),  80--92 (2006)

\bibitem{ferreira2022vtuber}
Ferreira, J.C.V., Regis, R.D.D., Gon{\c{c}}alves, P., Diniz, G.R., Tavares, V.P.d.S.C.: Vtuber concept review: The new frontier of virtual entertainment. In: Proceedings of the 24th Symposium on Virtual and Augmented Reality. pp. 83--96 (2022)

\bibitem{githubGitHubHecomiuLipSync}
hecomi: {G}it{H}ub - hecomi/u{L}ip{S}ync: {M}{F}{C}{C}-based {L}ip{S}ync plug-in for {U}nity using {J}ob {S}ystem and {B}urst {C}ompiler --- github.com. \url{https://github.com/hecomi/uLipSync} (2021), [Accessed 23-01-2025]

\bibitem{heeyoHeeyo}
Heeyo: {H}eeyo --- heeyo.life. \url{https://heeyo.life/} (2024), [Accessed 23-01-2025]

\bibitem{kasneci2023chatgpt}
Kasneci, E., Se{\ss}ler, K., K{\"u}chemann, S., Bannert, M., Dementieva, D., Fischer, F., Gasser, U., Groh, G., G{\"u}nnemann, S., H{\"u}llermeier, E., et~al.: Chatgpt for good? on opportunities and challenges of large language models for education. Learning and individual differences  \textbf{103},  102274 (2023)

\bibitem{Lin2024}
Lin, J., Chen, E., Gurung, A., Koedinger, K.R.: Mufin: A framework for automating multimodal feedback generation using generative artificial intelligence. In: Proceedings of the Eleventh ACM Conference on Learning@ Scale. pp. 550--552 (2024)

\bibitem{Makransky2018}
Makransky, G., Wismer, P., Mayer, R.E.: A gender matching effect in learning with pedagogical agents in an immersive virtual reality science simulation. Journal of Computer Assisted Learning  (Nov 2018). \doi{10.1111/jcal.12335}, \url{https://doi.org/10.1111/jcal.12335}

\bibitem{mcdonald2019reliability}
McDonald, N., Schoenebeck, S., Forte, A.: Reliability and inter-rater reliability in qualitative research: Norms and guidelines for cscw and hci practice. Proceedings of the ACM on human-computer interaction  \textbf{3}(CSCW),  1--23 (2019)

\bibitem{minn2022ai}
Minn, S.: Ai-assisted knowledge assessment techniques for adaptive learning environments. Computers and Education: Artificial Intelligence  \textbf{3},  100050 (2022)

\bibitem{okado2023can}
Okado, Y., Nye, B.D., Aguirre, A., Swartout, W.: Can virtual agents scale up mentoring?: insights from college students’ experiences using the careerfair. ai platform at an american hispanic-serving institution. In: International Conference on Artificial Intelligence in Education. pp. 189--201. Springer (2023)

\bibitem{self2016birth}
Self, J.: The birth of ijaied. International journal of artificial intelligence in education  \textbf{26}(1),  4--12 (2016)

\bibitem{sikstrom2024pedagogical}
Sikstr{\"o}m, P., Valentini, C., Sivunen, A., K{\"a}rkk{\"a}inen, T.: Pedagogical agents communicating and scaffolding students' learning: High school teachers' and students' perspectives. Computers \& Education  \textbf{222},  105140 (2024)

\bibitem{stamper2024enhancing}
Stamper, J., Xiao, R., Hou, X.: Enhancing llm-based feedback: Insights from intelligent tutoring systems and the learning sciences. In: International Conference on Artificial Intelligence in Education. pp. 32--43. Springer (2024)

\bibitem{Veletsianos2012}
Veletsianos, G.: How do learners respond to pedagogical agents that deliver social-oriented non-task messages? impact on student learning, perceptions, and experiences. Computers in Human Behavior  \textbf{28}(1),  275--283 (2012). \doi{10.1016/j.chb.2011.09.010}

\bibitem{wang2024large}
Wang, S., Xu, T., Li, H., Zhang, C., Liang, J., Tang, J., Yu, P.S., Wen, Q.: Large language models for education: A survey and outlook. arXiv preprint arXiv:2403.18105  (2024)

\bibitem{zhang2023sadtalker}
Zhang, W., Cun, X., Wang, X., Zhang, Y., Shen, X., Guo, Y., Shan, Y., Wang, F.: Sadtalker: Learning realistic 3d motion coefficients for stylized audio-driven single image talking face animation. In: Proceedings of the IEEE/CVF Conference on Computer Vision and Pattern Recognition. pp. 8652--8661 (2023)

\end{thebibliography}

\end{document}